\documentclass[%
 reprint,
superscriptaddress,
 amsmath,amssymb,
 aps,
]{revtex4-2}

\usepackage{graphicx}
\usepackage{dcolumn}
\usepackage{bm}
\usepackage[dvipsnames]{xcolor}
\usepackage[version=3]{mhchem}
\usepackage{chemarr}

\usepackage[dvipsnames]{xcolor}
\usepackage{bm}
\usepackage{comment}

\newcommand{\ldna}{$\lambda$-DNA~}

\begin{document}

\title{Effects of Monovalent and Divalent Cations on the Rheology of Entangled DNA}

\author{Jennifer Harnett}
\affiliation{School of Physics and Astronomy, University of Edinburgh, Peter Guthrie Tait Road, Edinburgh, EH9 3FD, UK}
\author{Simon Weir}
\affiliation{School of Physics and Astronomy, University of Edinburgh, Peter Guthrie Tait Road, Edinburgh, EH9 3FD, UK}
\author{Davide Michieletto}
\affiliation{School of Physics and Astronomy, University of Edinburgh, Peter Guthrie Tait Road, Edinburgh, EH9 3FD, UK}

\begin{abstract}
\textbf{In this paper we investigate the effects of varying cation valency and concentration on the rheology of entangled \ldna solutions. We show that monovalent cations moderately increase the viscoelasticty of the solutions mainly by stabilising linear concatenation of \ldna ``monomers'' via hybridisation of their sticky ends. On the contrary, divalent cations have a far more complex and dramatic effect on the rheology of the solution and we observe evidence of inter-molecular DNA-DNA bridging by Mg$^{2+}$. We argue that these results may be interesting in the context of dense solutions of single and double stranded DNA, e.g. in vivo or in biotechnology applications such as DNA origami and DNA hydrogels.}
\end{abstract}

\maketitle

\section{Introduction}

DNA is a charged anionic polyelectrolyte whose physical properties such as effective diameter~\cite{frank1997biophysics}, persistence length~\cite{lee2010influence, baumann1997ionic, hagerman1988flexibility} and twist ~\cite{cruz2022twisting} are influenced by both divalent and monovalent cations. Monovalent cations such as Na$^{+}$ and K$^{+}$, are abundant in cells~\cite{Alberts2014} and well known to screen electrostatic repulsion between the DNA phosphate groups and in general polyelectrolytes~\cite{Yang2022}. However, they are generally considered to not cause DNA-DNA attraction~\cite{qiu2006measuring, qiu2007inter}. Divalent cations, such as Mg$^{2+}$, also play essential roles in cells -- where they are typically present in mM range~\cite{Alberts2014,Hou2001} -- and are essential for some biological processes, for instance by facilitating interactions of proteins to DNA~\cite{farcaș2020influence} or signalling between cells~\cite{berridge2000versatility}. Moreover, \emph{In vitro}, cations and especially MgCl$_2$ are key to self-assemble and stabilise DNA origami structures~\cite{castro2011primer}, as well as commonly being used to absorb DNA onto negatively charged mica for AFM studies~\cite{main2021atomic, pastre2003adsorption}. 

Because of the widespread presence of cations \emph{in vivo} and \emph{in vitro}, their effects on DNA-DNA interactions need to be well understood. The condensation and phase separation of DNA in the presence of cations with a valency Z$\geq$ 3, e.g. spermidine, spermine, and cobalt hexammine, has been extensively studied and shown in magnetic tweezers experiments~\cite{bloomfield1997dna, widom1983monomolecular, sun2019multiscale,todd2008interplay}. In a bulk solution, no condensation of double-stranded DNA has been observed in the presence of only divalent cations, even up to concentrations of 1M MgCl$_2$~\cite{koltover2000dna}. However, cations-induced condensation has been observed in certain specific conditions. For instance, triple-stranded DNA undergoes condensation at concentrations as low as 10 mM MgCl$_2$~\cite{qiu2010divalent, zhang2017divalent}. Furthermore, the structure of DNA grooves (and hence DNA sequence) has been found to play a crucial role in DNA-DNA interactions. For instance, alkaline earth metals have been shown to condense ``AATT'' repeating sequences~\cite{srivastava2020structure}. Moreover, confinement and alignment of DNA molecules influence their behavior in the presence of divalent cations, as demonstrated by DNA condensation when confined in 2D on a cationic surface~\cite{koltover2000dna}. 

Albeit not causing condensation of DNA in aqueous solutions, divalent cations such as MgCl$_2$ can induce ion bridging in DNA~\cite{zhang2017divalent}. Theoretical predictions and simulations~\cite{tan2006ion, zhang2017divalent, luan2008dna} have proposed the presence of an attractive force between DNA strands in the presence of divalent salts. However, the experimental evidence of this remain limited. 
For instance, X-ray scattering suggests the presence of an effective short-range attraction between short DNA molecules at concentrations as low as 16 mM MgCl$_2$~\cite{qiu2006measuring} but that this attraction weakened with increasing DNA length~\cite{qiu2007inter}. 


The largest majority of the work done on understanding the cation-mediated interaction between DNA molecules considered dilute conditions or even single-molecule setups~\cite{qiu2006measuring, qiu2007inter, zhang2017divalent}. However, in many situations cations are affecting the behaviour of DNA at high concentrations, for instance in cells where DNA volume fraction is $>$ 2\%, in the delivery of DNA origami cargos ~\cite{balakrishnan2019delivering} where one would require high concentration payload to be effective and, finally, in RNA and DNA vaccines as they are injected into the bodies at high concentrations~\cite{gary2020dna}. Additionally, variations in salt concentration, which influence DNA-DNA interactions, will likely impact the rheological properties of DNA-hydrogels ~\cite{conrad2019increasing, biffi2013phase, peng2022dynamic, xing2018microrheology}. In spite of this, there is still limited understanding on the consequence of salt valency and concentration on the rheology of entangled DNA solutions.   

\begin{figure*}  
\centering
\includegraphics[width=\textwidth]{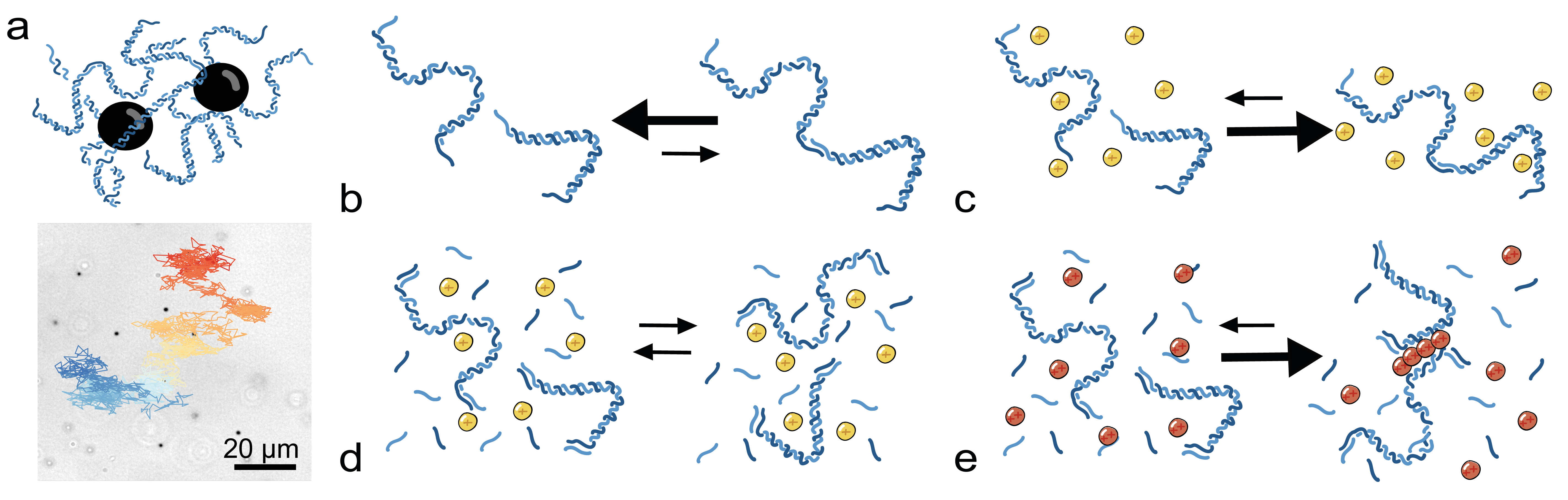}
\caption{(\textbf{a}) We employ microrheology~\cite{Mason1995} to quantify the behaviour of dense solutions of \ldna under different salt types and concentrations. Microrheology is done by recording movies of tracer beads (see bottom panel) diffusing in a fluid. In the panel we also superpose the trajectory of a single bead, where increasing time is schematised from red to blue. (\textbf{b}) \ldna undergoes salt dependent hybridisation through its ``sticky'' ends. (\textbf{c}) Monovalent cations stabilise the Watson-Crick base pairing and favour the formation of longer \ldna polymers. (\textbf{d}) Introducing super-stoichiometric quantities of short ssDNA oligos quenches the hybridisation of \ldna sticky ends. (\textbf{e}) Using this assay, we can investigate if in the presence of ssDNA oligos, divalent cations may still affect the solution rheology via DNA-DNA bridging. }
\label{fig:intro}  
\end{figure*}

To address this gap, in this paper we investigate the effects of cation concentration and valency on the rheology of entangled DNA solutions (see Fig.~\ref{fig:intro}). More specifically, we consider \ldna as it is a highly monodisperse polymer~\cite{Banik2021} that displays two ``sticky'' ends with 12 unpaired nucleotides enabling concatenation via hybridisation~\cite{becker1990bacteriophage,Haber2000}. The hybridisation and melting reactions of the sticky ends are sensitive to the salts in solution (as well as the temperature), and can be thought of as akin to the fusion and breakage of worm-like micelles~\cite{cates1987reptation, Michieletto2022natcomm}. Thus, we expect a distribution of concatenamer lengths that depends on the salt concentration and valency and, in turn, an effect on the rheology of the solution. Indeed, a key result of our paper is that at large concentrations of divalent cation MgCl$_2$ (50 mM), solutions of \ldna increase their viscosity up to 43-fold, with respects to the case with no cations. On the other hand, even in presence of 0.5 M monovalent cation NaCl, we observe a more modest 7-fold increase.  

Beyond the effect on the sticky ends hybridisation, we also expect the rheology of entangled \ldna to be sensitive to the DNA-DNA cation-mediated interactions. We test this hypothesis by making use of short single-stranded DNA oligomers (``oligos'') that can quench the hybridisation of \ldna sticky ends, enabling us to readily turn off the concatenation process. In this set up, we discover that even in the presence of super-stoichiometric quantities of oligos, solutions of \ldna with increasing MgCl$_2$ display a slowing down which we conclude must be linked with inter-molecular bridging by divalent cations. 

We argue that these results ought to be relevant to better understand mobility, dynamics and rheology of DNA \emph{in vivo}, with potential consequences on the delivery of highly concentrated nucleid acids for instance in DNA/RNA vaccines and DNA hydrogels. 

\section{Methods}
\subsection{\ldna Preparation}
\ldna was purchased from New England Biolab at a concentration of 500 ng/$\mu$L ($\simeq 16$ nM of \ldna molecules) in TE buffer (10 mM Tris-HCl pH 8, 1 mM EDTA) and stored at -20$^{\circ}$C. The \ldna with sticky ends (no oligos) is used directly from the stock solution. The 12 nucleotides ssDNA oligos are purchased from IDT with sequences ($\lambda_1=$ 5’ GGGCGGCGACCT 3’ and $\lambda_2$=5’ AGGTCGCCGCCC 3') and resuspended in TE buffer at concentration of 100 $\mu$M. To prepare samples of \ldna with oligos we mix 98 $\mu$L of \ldna stock with 1 $\mu$L of oligo $\lambda_1$ and 1 $\mu$L of oligo $\lambda_2$. The stoichiometry is thus set to around 60 oligos per sticky end. After mixing, we heat the solutions (both with and without oligos) at 65$^{\circ}$C for 10 minutes, mixed by pipetting with wide bore tips and finally let them cool at room temperature. Given the tendency of \ldna solutions to display heterogeneous behaviour~\cite{Banik2021}, we leave our solutions on a roller bank to homogenise for at least 24 hours before use. 

\subsection{Microrheology}

\begin{figure}[t!]
\centering
\includegraphics[width=0.5\textwidth]{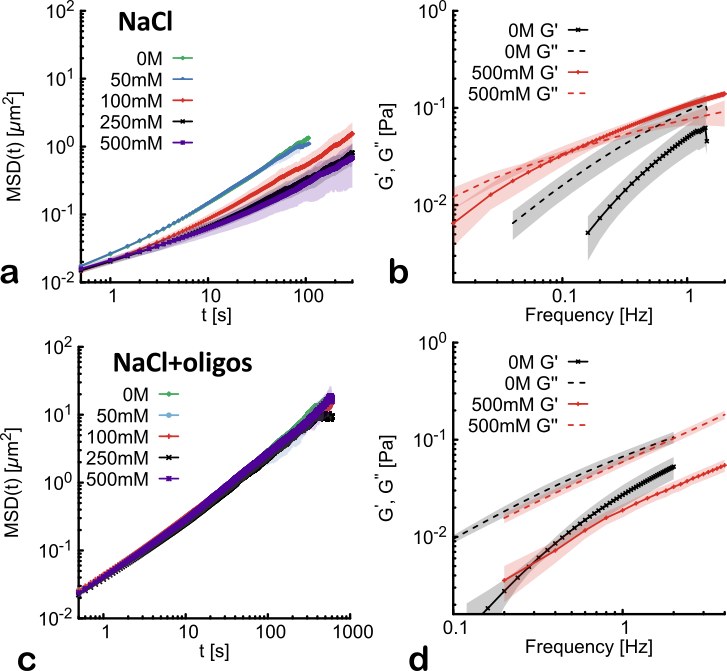}
\caption{ (\textbf{a}) Mean squared displacement (MSD) as a function of lag time ($t$) with increasing concentrations of monovalent cations (NaCl).The shaded area of the curves represents the standard deviation with respect to positions in the same sample and across samples.(\textbf{b}) Elastic ($G^{'}$) and viscous ($G^{''}$) moduli obtained via apply the  Generalised Stokes-Einstein Relation (GSER) for the case 500 mM NaCl. (\textbf{c}) MSD of the same systems as in (\textbf{a}) in the presence of super-stoichiometric quantities of ssDNA oligo quenchers.(\textbf{d}) $G^{'}$ and $G^{''}$ obtained from the MSD at 500 mM NaCl and in presence of ssDNA oligos. The shaded area of the $G^{'}$ and $G^{''}$ curves represents the standard deviation between different video repeats.}
\label{fig:NaCl_MSD} 
\end{figure}

After the homogenisation step, 9 $\mu$L of 500 ng/$\mu$L \ldna was mixed with 1/$\mu$L of 10X salt buffer to create a solution of approximately 450 ng/$\mu$L \ldna at the desired salt concentration.  The concentration of DNA was checked on a Nanodrop before each experiment and determined to be 450 $\pm$ 25 ng/$\mu$L for all the experiments reported here, and a gel electrophoresis was run to check the integrity of \ldna regularly. 
We recall that the overlap concentration of \ldna is $c^* \simeq 20$ ng/$\mu$L, i.e. around 25 times smaller than the concentrations considered in this work. Through analysis of the probability distribution functions (PDFs) of the particle displacements, we discovered that the samples were non-homogeneous after the addition of cations (see SI Figure 3). For this reason, all samples were left on a roller bank at room temperature to equilibrate for at least 24hrs in presence of cations, resulting in PDFs of particle displacements which could be well fitted by simple Gaussians, suggesting that the fluids were at that point homogeneous (SI Figure 4). After this equilibration step, we also checked that the samples did not show degradation by performing gel electrophoresis of small aliquots. Polystyrene microspheres of diameter $a = 1.1$ $\mu$m were coated with BSA to prevent aggregation under high ionic strength buffers as well as binding interactions with the DNA. The beads were vortexed prior to use and 0.4$\mu$L added to 10$\mu$L DNA solution and dispersed uniformly throughout the sample by mixing using cut tips. The sample chamber was created by attaching a double sided sticky tape (100 $\mu$m thick) on a microscope slide. In each square 5 $\mu$l of sample was loaded, with two replicas per sample being performed. Finally, the chamber was sealed with a glass coverslip. The particles were visualised using a Nikon Eclipse Ts2 microscope with a 60x objective. Videos were recorded at a speed of 2 fps on a 1024x1024 ROI (approximately 30 particles in the ROI). We recorded 10 different positions in each sample, resulting in around 300 independent tracks per sample. The movies are recorded using an exposure of 4ms and intervals of 0.5 seconds and for up to 1000 seconds, to span a broad range of timescales. All microrheology experiments were performed in a temperature-controlled Okolab chamber stage-top incubator at 25$^{\circ}$C. 

Particle tracking was done using TrackPy and custom-written particle-tracking codes (in Python and C++). From the trajectories we measured the time averaged mean squared displacement (MSD) of the particles as a function of lag time $t$ as $MSD(t) = \langle \left[ r(t_0) - r(t+t_0)\right]^2 \rangle$ where the average is performed over particles and times $t_0$ and $r$ is either the x or y component of the 2D position $\vec{r}$. We corrected for static error~\cite{savin2005static} by taking videos of tracer probes fixed to the glass substrate and removing their MSD from the one of the probes in the bulk; however, the contribution of static error is negligible as the MSD of the stuck probes are at least 2 orders of magnitude smaller than the MSD of the particles in the bulk. We also corrected for dynamic errors by taking different videos at varied exposure times.  We observed that the MSD of the probes were affected for exposure times longer than 50 ms, thus all the videos were taken at 4 ms exposure time.  

The averaged MSD along $x$ and $y$ directions was then used to extract the diffusion coefficient $D$ of the beads as $D = \lim_{t \to \infty} MSD(t)/2t$. In practice, we fitted the MSD at the largest timelags in the trajectories and computed it using 3 different lagtime ranges and then averaged. The viscosity of each sample is then calculated using the Stokes-Einstein equation $\eta = k_B T/(3 \pi D a)$.

The viscous and elastic moduli are calculated by using the generalised Stokes-Einstein relation (GSER)~\cite{Mason2000}. Briefly, we fitted the MSD using a polynomial function and we then extracted the complex modulus as
\begin{equation}
    |G^{*}(\omega)| = \dfrac{k_B T}{3 \pi a MSD(1/\omega) \Gamma[ 1 + \alpha(\omega)]}
\end{equation}
where $\alpha(\omega) = \left. d \log{MSD(t)} / d \log{t} \right|_{t = 1/\omega}$ is the MSD exponent as a function of lagtime and $\Gamma$ is the Gamma function. The viscous $G^{''}$ and elastic $G^{'}$ moduli are then computed as 
\begin{align}
    & G^{'}\omega) = |G^{*}(\omega)| \sin{(\pi \alpha(\omega)/2)} \\
    & G^{''}(\omega) = |G^{*}(\omega)| \cos{(\pi \alpha(\omega)/2)} \, .
\end{align}
In practice, we apply the GSER to at least three independent samples MSD, and report the the mean values of $G^{'}$ and $G^{''}$ and their standard deviations.

\subsection{Free Energy Analysis}
The free energy of the secondary structures of the hybridised sticky ends were calculated using NUPACK~\cite{nupack} (\url{https://www.nupack.org}). We entered the specific 12 bp sequence of \ldna sticky end as well as NaCl and MgCl$_2$ concentration. The software computes basepair, stacking and cross-stacking energies and returns a hybridisation free energy of the secondary structure. We note that NUPACK requires that a minimum of 50 mM NaCl is present, and therefore all free energies as a function of MgCl$_2$ concentration were obtained including 50 mM NaCl in the software settings. 
We also checked that the trends of the free energies as a function of cation concentration reported in this paper were in agreement with the ones extracted from a different software, DINAMelt \cite{markham2005dinamelt}(\url{http://www.unafold.org}). 


\section{Results}
\subsection{Monovalent cations}

We first investigate how increasing the concentration of \emph{monovalent} cations in entangled solutions of \ldna affects the fluid's rheology. Figure~\ref{fig:NaCl_MSD} shows the viscoelastic behaviour of fluids with increasing concentrations of NaCl between 0 and 500 mM. In absence of quencher oligos (Fig.~\ref{fig:NaCl_MSD}a) we observe that the passive tracers embedded in the fluid display slower diffusion with increasing [NaCl]. We also note the onset of an elastic behaviour at short timescales for [NaCl]$ > 100 $ mM, as captured by the subdiffusive behaviour of the MSDs at lagtimes $t < 10$ seconds. To best quantify the onset of elasticity we use the Generalised Stokes-Einstein Relation~\cite{Mason1995,Mason2000} to obtain $G^{'}$ and $G^{''}$. For the case with 500 mM NaCl, we clearly observe that the elastic modulus dominates over the viscous modulus at large frequencies $\omega > 0.1$ $s^{-1}$ (Fig.~\ref{fig:NaCl_MSD}b). Interestingly, the observed slowing down due to increasing [NaCl] displays a significant increase around 100 mM. This is in line with results obtained in dilute conditions of short DNA duplexes, as at this value of [NaCl] a significant screening of electrostatic repulsion was measured in SAXS~\cite{qiu2006measuring}. We also find that this slowing down appears to plateau around 250-500 mM NaCl. 

\begin{figure}[t!]
\centering
\includegraphics[width=0.5\textwidth]{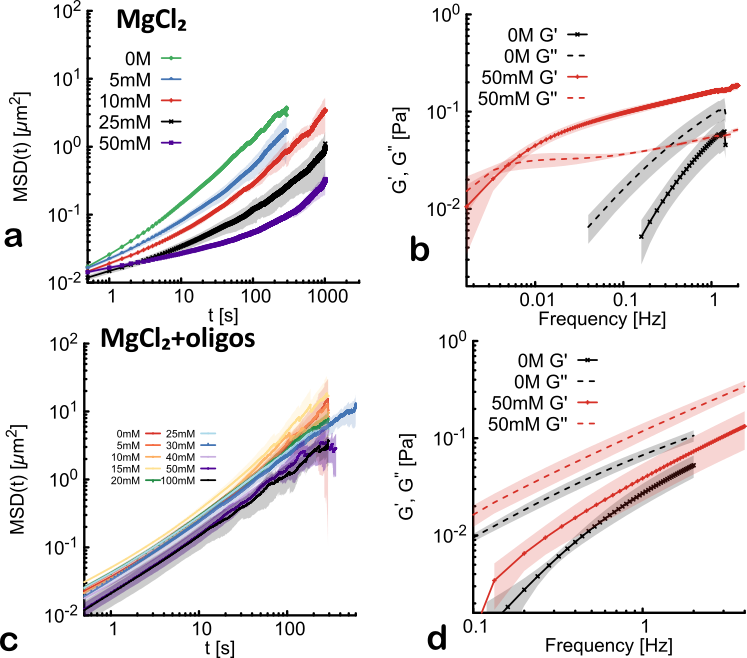}
\caption{(\textbf{a}) Mean squared displacement (MSD) as a function of lag time ($t$) with increasing concentrations of divalent cations (MgCl$_2$).The shaded area of the curves represents the standard deviation with respect to positions in the same sample and across samples. (\textbf{b}) Elastic ($G^{'}$) and viscous ($G^{''}$) moduli obtained via the GSER for the case 50 mM MgCl$_2$.(\textbf{c}) MSD of the same systems as in (\textbf{a}) in the presence of super-stoichiometric quantities of ssDNA oligo quenchers.  (\textbf{d}) $G^{'}$ and $G^{''}$ obtained from the MSD at 50 mM MgCl$_2$ and in presence of ssDNA oligos.The shaded area of the $G^{'}$ and $G^{''}$ curves represents the standard deviation between different video repeats.}
\label{fig:MgCl2_MSD}
\end{figure}

In marked contrast, we observe a completely different behavior when we introduce super-stoichiometric quantities of ssDNA oligos that quench the hybridisation of the \ldna sticky ends. Indeed, in this case we observe no increase in viscosity or onset of elasticity for any value of [NaCl] considered in this work, up to 0.5M NaCl (Fig.~\ref{fig:NaCl_MSD}c-d). Throughout the concentration range, the MSDs appear remarkably insensitive to [NaCl]. This finding may appear at odds with the fact that NaCl screens DNA self-interactions and reduces its persistence length, effectively rendering individual \ldna coils more compact~\cite{Rybenkov1993}. While this is certainly true, we note that our solutions are well above overlap concentration (which for \ldna is $c^* \simeq 20$ ng/$\mu$L), entailing that NaCl will mostly screen inter-molecular interactions and leave the single chains unaffected. For this reason we expect that even at large [NaCl] the \ldna coils will remain well entangled with each other and so display [NaCl]-insensitive viscoelastic behaviours in absence of sticky ends. Our microrheology supports this expectation, as we find no change in the passive tracers' behaviour over a wide range of [NaCl]. 

In conclusion, our experiments support the hypothesis that monovalent cation NaCl stabilises sticky ends mediated concatenation of \ldna molecules, in turn triggering elasticity and driving an increase in viscosity. On the contrary, in the absence of sticky ends, our data support a model whereby there is no net effect of NaCl on the rheology of the DNA solution, even at concentrations as large as 0.5 M. 

\subsection{Divalent cations}

Divalent cations are often used to stabilise DNA origami structures~\cite{castro2011primer} or absorb DNA on mica~\cite{Lyubchenko2002} and have been found to effectively change the interaction potential from repulsive to attractive between DNA molecules~\cite{qiu2006measuring}. Perhaps more importantly, Mg$^{2+}$ is essential to allow ATP (or CTP) binding and hydrolysis in certain protein complexes, such as condensin~\cite{Shaltiel2022} and parB~\cite{Tisma2022}, which act on the highly crowded and entangled genome. Thus, we decided to investigate what is the effect on the rheology of dense solutions of \ldna at varying [MgCl$_2$]. It is clear from Fig.~\ref{fig:MgCl2_MSD}a that divalent cations have a far stronger effect on the rheology of the solutions; indeed, the beads display a persistent subdiffusive behaviour already at [MgCl$_2$] $\simeq  10$ mM. Overall, the change in rheological properties is far more pronounced when increasing the concentration of divalent cations compared to monovalent cations, with a large reduction in the diffusivity of the beads between 0 mM and 50 mM MgCl$_2$. We also note that for 50 mM MgCl$_2$, the elasticity-dominated regime appears at much lower frequencies than in the case of NaCl (around $\omega \simeq 10^{-2} s^{-1}$, see Fig.~\ref{fig:MgCl2_MSD}b), implying a far longer relaxation time. On the contrary, the value of the elastic modulus at larger frequencies, say $G^{'}(1 Hz)$, is similar for both salts (around 0.1 Pa). In light of this, we thus argue that the main contribution of MgCl$_2$ to the fluid viscoelasticity is to create more stable \ldna concatenamers, which are longer lived than the ones in the presence of NaCl.

To verify the role of the sticky ends, we again repeat the same experiment in presence of super-stoichiometric quantities of ssDNA oligo quenchers. Interestingly, and in marked contrast with the previously seen insensitivity of quenched \ldna in increasing NaCl, we here observe a significant, albeit moderate, slowing down (Fig.~\ref{fig:MgCl2_MSD}c). In fact, the behaviour of the MSDs of the tracer beads is qualitatively different from the case without oligos: they appear to slow down but there is no onset of subdiffusion within our experimental timscales. The absence of elasticity-dominated regime is also confirmed through the GSER, whereby we do not observe any crossover point between G$^{'}$ and G$^{''}$ (Fig.~\ref{fig:MgCl2_MSD}d).

It is interesting to note that the slowing down of the beads' MSD appears most notable for [MgCl$_2$] greater than 30mM (Fig.~\ref{fig:MgCl2_MSD}c), a regime in which SAXS data measured an effective attraction between short dsDNA molecules~\cite{qiu2006measuring}. This suggests that the onset of slowing down in  solutions of \ldna where the sticky ends are quenched may be due to DNA-DNA interactions mediated by divalent cations, forming transient bridges between the molecules (see also Fig.~\ref{fig:intro}e).


\subsection{Viscosity dependence on cation concentration and valency}

A summary of our results can be found in Fig.~\ref{fig:ETA}, where we show the behaviour of the normalised viscosity of the system $\eta/\eta_0$ (obtained via the Stokes-Einstein relation and where $\eta_0$ is the value of viscosity for the case with no salt) as a function of salt concentration. One can readily notice that NaCl induces a weaker increase, which is absent in the quenched \ldna solution, whereas the case with MgCl$_2$ displays (i) a steeper increase in viscosity and (ii) a moderate viscosity increase even in the quenched case.

As seen before, samples with increasing NaCl display a moderate increase in viscosity and a plateau around $\eta/\eta_0 \simeq 7 \pm 2$, while adding divalent cations have a far bigger effect, with a $\eta/\eta_0 \simeq 43 \pm 7$ increase in viscosity observed between 0 and 50 mM MgCl$_2$. On the contrary, we can clearly appreciate that the viscosity remains approximately constant with increasing NaCl when the oligomers are added to the solution, rendering them ``quenched''. Again, this suggests that the rheology is mainly dictated by the ability of sticky ends to hybridise and form effectively longer polymers by concatenating \ldna together~\cite{Rubinsteinbook}.  
In contrast, adding divalent cations has (i) a much steeper increase in viscosity and (ii) an effect on viscosity even in the presence of sticky ends quenchers.

\subsection{Analogy with Worm-like Micelles and Scaling of Viscosity with Cations}

To better understand how the viscosity of the solution may scale with salt concentration, we draw an analogy with systems of worm-like micelles, which can undergo breakage and fusion in equilibrium~\cite{Cates2006a,Michieletto2020}. In our system, single \ldna ``monomers'' can fuse and break through sticky ends hybridisation and melting to create an equilibrium with a characteristic polymer contour length~\cite{cates1987reptation}
\begin{equation}
\Bar{L} \sim \left( \dfrac{c_2}{c_1} \right)^{1/2}
\end{equation}
where $c_2$ is the rate of fusion and $c_1$ the rate of breakage. In our system, both the rate of hybridisation and that of melting depend on the salt concentration and valency~\cite{SantaLucia1996,SantaLucia1998,Banerjee2021,Tan2006}. For a two-state model 
\begin{equation}
    A+A  \xrightleftharpoons[c_1]{c_2}  2A
\end{equation}
the equilibrium constant is related to the free energy $\Delta G$ of the secondary structure as $k_D = c_2/c_1 = \exp{(- \Delta G/k_BT)}$. 

\begin{figure}[t!]
\centering
\includegraphics[width=0.45\textwidth]{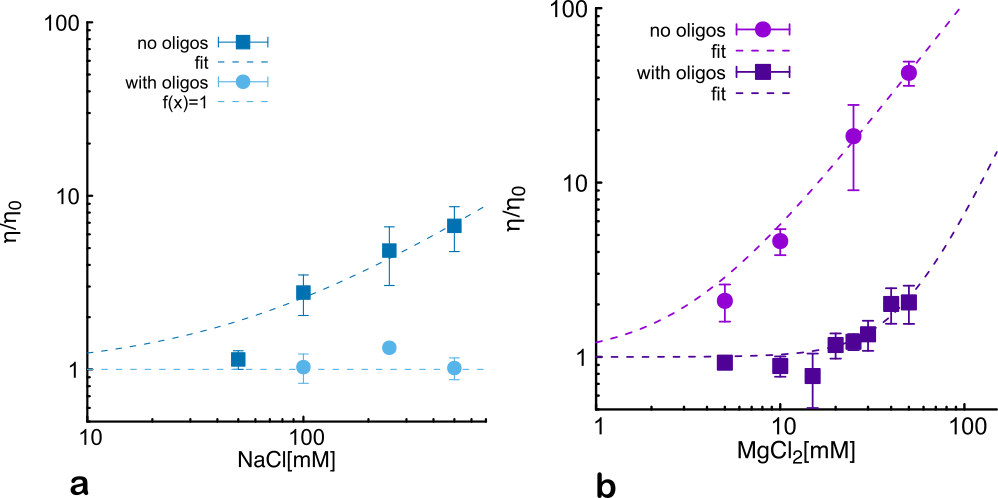}
\caption{Viscosity as a function of cation concentration for both monovalent (a) and divalent cations (b). Viscosity values $\eta$ are normalised with respect to the viscosity of \ldna at 450ng/$\mu$L with no cations present ($\eta_0$). Both \ldna with ssDNA oligos (squares) and without oligos (circles) are plotted together. Points correspond to the experimental data with errors, and the dashed lines display a power law fitted to the data in the form $f(x) = 1+Ax^{b}$. We find the best fit parameters when no oligos are present are as follows: $A_{NaCl}= 0.037 \pm 0.041$ , $b_{NaCl} = 0.82 \pm 0.19$ for solutions with NaCl, and  $A_{MgCl_2}=0.21\pm 0.07 $ , $b_{MgCl_2} = 1.35 \pm 0.09$ for solutions with MgCl$_2$. When oligos are present for monovalent cations we plot f(x)=1, and for divalent cations we again fit a power law with best parameters: $A_{MgCl_2}=0.00020 \pm 0.00006 $ , $b_{MgCl_2} = 2.24 \pm 0.07$}
\label{fig:ETA}      
\end{figure}

\begin{figure}[t!]
\centering
\includegraphics[width=0.45\textwidth]{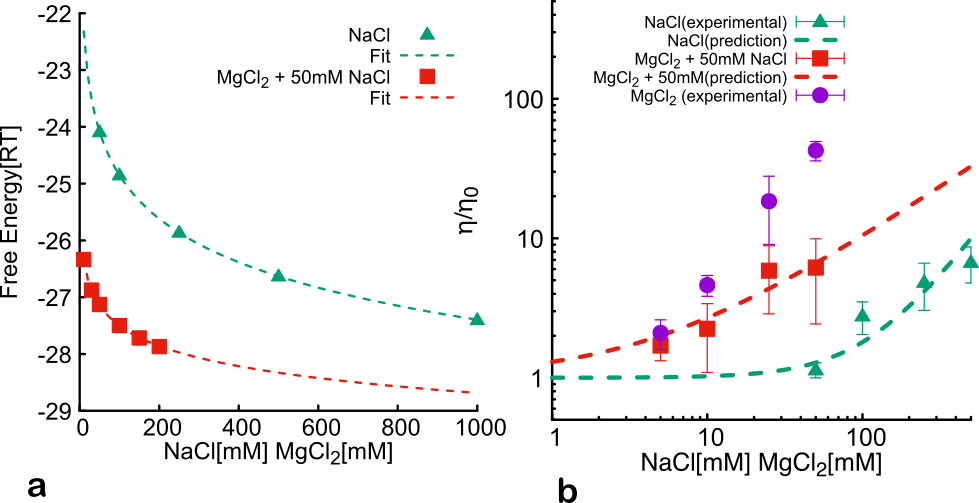}
\caption{(a)The free energy in units RT as a function of cation concentration. NaCl is shown in green and denoted by triangles, MgCl$_2$ (with 50mM NaCl for all values) is shown in red and denoted by squares.  Points correspond to the Nupack data \cite{nupack} and the dashed lines display a logarithmic function fitted to the data in the form \textit{$f_s(x) =  a_s  - b_s \log({x})$}. The best fit parameters are as follows: $a_{NaCl} = -19.763 \pm 0.022 $, $b_{NaCl} = -1.105 \pm 0.004$, $a_{MgCl2} = -25.138 \pm 0.061$, $b_{MgCl_2} = -0.514 \pm 0.014$. (b) Dependence of the normalised viscosity on cation concentration for NaCl (green triangles), MgCl$_2$ (purple circles) and MgCl$_2$+50mM NaCl (red squares). The power law $\eta_{NaCl} \sim x^{3/2}$ is plotted in green dashed line and $\eta_{MgCl2} \sim x^{3/4}$ in red dashed line to compare to the experimental results. }  
\label{fig:ETA_scaling}              
\end{figure}

To quantify how $k_D = c_2/c_1$ scales with salt concentration, we compute the salt-dependent free energy $\Delta G$ using NUPACK~\cite{nupack} (and similar results were obtained with DINAMelt). 
In Fig.~\ref{fig:ETA_scaling}a, we show $\Delta G$ as a function of salt concentration for NaCl and MgCl$_2$. Motivated by the tightly bound ion (TBI) model -- predicting a logarithmic dependence of the duplex free energy on salt concentration~\cite{Springer2010,Tan2006} -- we fit these points with the function 
\begin{equation}
   f_s(x) =  a_s  + b_s \log({x})
   \label{eq:tbi}
\end{equation}
which returns a good fit with parameters $a_{NaCl} = -19.76 \pm 0.02 $, $b_{NaCl} = -1.105 \pm 0.004$, $a_{MgCl2} = -25.14 \pm 0.06$, $b_{MgCl2} = -0.51 \pm 0.01$. We can therefore include this result to estimate the characteristic length as 
\begin{equation}
\Bar{L} \sim \left(e^{- a_s - b_s \log{x}} \right)^{1/2} \sim x^{- b_s/2}
\end{equation}
where we have discarded constants that are salt independent, and where $x$ indicates salt concentration. In the entangled regime, we thus expect the viscosity to scale as~\cite{Doi1988} 
\begin{equation}
    \eta_s \sim \bar{L}^3 \sim x^{-3 b_s/2}
\end{equation}
and more specifically, we expect $\eta_{NaCl} \sim x^{3/2}$ and $\eta_{MgCl2} \sim x^{3/4}$. To test these predictions we fitted the values of $\eta/\eta_0$ in Fig.~\ref{fig:ETA} and obtained $\eta/\eta_0 \sim x^{0.82 \pm 0.19}$ for solutions with NaCl, and $\eta/\eta_0 \sim x^{1.35 \pm 0.09}$ for solutions with MgCl$_2$. However, when the NaCl data is fitted between 0 mM - 250 mM (before the plateau) we find $b_{NaCl} = 1.14 \pm 0.29$ which is in better agreement with the prediction. We find the influence of MgCl$_2$ concentration on the viscosity is far greater than the predicted trend. We argue that the discrepancy with the MgCl$_2$ data may be due to the short-range attraction of dsDNA molecules induced by Mg$^{+2}$ ions that is not accounted for by NUPACK and other secondary structure stability algorithm. 

However, motivated by the fact that the predicted values of hybridisation free energy always include 50 mM NaCl when calculated in NUPACK, we then also investigated how varying MgCl$_2$ concentration in a background of 50 mM NaCl would affect the viscosity. In Figure~\ref{fig:ETA_scaling}b we show that the steep viscosity increase is attenuated, and is reduced 8-fold by the presence of 50 mM NaCl. This suggests that NaCl competes with MgCl$_2$ to interact with DNA. This finding is also in line with the fact that at this ratio of NaCl to MgCl$_2$ it is predicted that NaCl will dominate (at least in the case of short DNA duplexes)~\cite{owczarzy2008predicting}. Furthermore, the data was again fitted to a power law function $f(x) = 1+Ax^{b}$ (same as Figure~\ref{fig:ETA}) and we found $a_{MgCl2+50mM NaCl} = 0.4 \pm 0.1$, $b_{MgCl2+50mM NaCl} = 0.7 \pm 0.1$ which is in better agreement with the one predicted using the TBI model ($\eta_{MgCl2}\sim x^{3/4}$). 

\section{Conclusions}
In conclusion, in this paper we investigated the broadly overlooked effect of cation valency and concentration on the rheology of a solution of \ldna. Using microrheology we discovered that increasing monovalent cation concentration leads to a reduction in the tracers' mobility, indicating a rise in viscosity, along with an onset of elastic behaviour at short time scales. Interestingly, the introduction of oligomers quenching the \ldna sticky ends removes this dependence on cation concentration and suppresses the onset of elasticity. This suggests that the changes in viscoelasticity of the DNA solution are mainly due to the cation-mediated increase in stability of sticky ends hybridisation, in turn allowing longer \ldna concatenamers (see Fig.~\ref{fig:NaCl_MSD}). 
 
On the other hand, we discover that divalent cations display a stronger correlation between concentration and viscoelasticity, characterized by notably longer relaxation times. Unlike with monovalent cations, the presence of quenching oligomers did not completely remove the cation-induced thickening, although the observed increase in elastic behavior was absent. This suggests that divalent cations induce attractive DNA-DNA interactions beyond stabilising the hybridisation of the sticky ends (see Fig~\ref{fig:MgCl2_MSD}).

Our findings on the viscosity as a function of cation concentration suggest power-law relationships: for monovalent cations, the viscosity followed  $\eta/\eta_0 \sim x^{0.82 \pm 0.19}$ while for divalent cations, it displayed a more pronounced scaling $\eta/\eta_0 \sim x^{1.35 \pm 0.09}$ (see Fig.~\ref{fig:ETA}). These results are in broad agreement with the prediction from tightly bound ion model coupled with that from worm-like micelles, describing the average length of concatenamers as a function of scission and fusion kinetics~\cite{cates1987reptation}. Using these theories, we find a dependence of $\eta_{NaCl}\sim x^{3/2}$ for monovalent cations and $\eta_{MgCl2}\sim x^{3/4}$ for divalent cations in a background of 50 mM NaCl (via NUPACK free energy calcuation), and we find that these predictions hold in the range 0 - 250 mM NaCl. Importantly, when monovalent and divalent cations were combined, the dependence of viscosity on cation concentration underwent a significant reduction, declining from $\eta/\eta_0 \sim x^{1.35 \pm 0.09}$ to $\eta/\eta_0 \sim x^{0.68 \pm 0.08}$ when 50mM NaCl was added to MgCl$_2$, then agreeing with the prediction within the uncertainty. This underscores the intricate interplay between different cation types and their influence on the viscoelastic behavior of DNA solutions (Fig.~\ref{fig:ETA_scaling}).

Overall, our study contributes to a better understanding of how cations affect the rheology of DNA solutions and offers a simple read-out to determine cation-mediated attractive DNA-DNA interactions, which remain elusive to firmly quantify~\cite{qiu2007inter}. We aim to extend this approach in the future to multivalent cations, such as spermine, and hydrotropic salts such as NaSal and ATP.

\section*{Author Contributions}
JH performed experiments and data analysis. DM supervised the project. 
All authors contributed to writing the paper.

\section*{Conflicts of interest}
There are no conflicts to declare

\section*{Acknowledgements}
DM thanks the Royal Society for support through a University Research Fellowship. This project has received funding from the European Research Council (ERC) under the European Union's Horizon 2020 research and innovation program (grant agreement No 947918, TAP) and through Soft Matter for Formulation and Industrial Innovation (SOFI CDT) (grant reference EP/S023631/1). We thank Andreia Fonseca Da Silva for her help with rheology experiments. For the purpose of open access, the author has applied a Creative Commons Attribution (CC BY) licence to any Author Accepted Manuscript version arising from this submission.

\bibliography{rsc} 
\bibliographystyle{apsrev4-1} 

\end{document}